\documentstyle[preprint,aps]{revtex}
\begin {document}
\draft
\preprint{UCI-TR 96-31}
\title{Approach to Microcanonical Equilibrium for Nonisolated Systems}
\author{Myron Bander\footnote{Electronic address: 
mbander@funth.ps.uci.edu}}
\address{ Department of Physics, University of California, Irvine, 
California
92657-4575} 

\date{September, 1996}
\maketitle
\begin{abstract}
The approach to equilibrium for systems interacting with their
environment by being regularly exposed to low energy, low intensity
pulses of some type of quanta is studied. Assuming a randomness
condition on the interaction of these quanta with the system, but
making no assumptions about the accessible states of the system, we
show that microcanonical equilibrium is reached. Although the
intensity of the pulses is assumed to be weak the interactions are
treated exactly, with no recourse to perturbation theory. The origin
of time asymmetry and the absence of recurrence is discussed.

\end{abstract}


\section{Introduction}\label{Introduction}  
For over a 100 years various approaches have been used to study the
foundations of statistical mechanics. That this subject continues to
be of interest \cite{Lebowitz} indicates that no completely
satisfactory description exists; this is true, both, for classical and
quantum statistical mechanics. The question of how macroscopic
irreversibility arises from reversible microscopic dynamics continues
to be of a topic of discussion and is reviewed in
Ref.~\cite{Lebowitz}.

For quantum statistical mechanics the goal to show that a system can
be described by a microcanonical density matrix, which for some given
energy $E$ is
\begin{equation}\label{densitymatrix}
\mbox{\boldmath $\rho$}_{i,j}(E)=\frac{\delta_{i,j}}{N(E)}\, ,
\end{equation} 
where $N(E)$ is the number of degenerate eigenstates with energy
$E$. It would be desirable to obtain the completely mixed state
represented by the density matrix of eq.~(\ref{densitymatrix}) as a
time development of a pure initial state; this is of course impossible
under unitary time evolution and one has to resort to various
averaging assumptions. In the most straight forward ensemble
``derivation'' of this result one essentially assumes the answer by
postulating random phases and equal {\it \`{a} priori} probabilities
for the eigenstates of some Hamiltonian.  More recently developments
in the theory of chaotic systems have lead to a different ways of
obtaining quantum statistical mechanics \cite{chaos}.  An old approach
to  (\ref{densitymatrix}) is through the Pauli master equation
\cite{Cohen},
\begin{equation}\label{Pauli}
\frac{d{\cal P}_i(t)}{dt}=\sum_n\left[t_{i\leftarrow n}{\cal P}_n(t)
   -t_{n\leftarrow i}{\cal P}_i(t)\right]\, ,
\end{equation} 
where ${\cal P}_i(t)$ is the probability at time $t$ of the system
being in the state $i$ and $t_{i\leftarrow n}$ is the rate per unit
time to go from state $n$ to state $i$. In order to obtain
eq.~(\ref{Pauli}) certain assumptions have to hold\cite{Cohen}:
\begin{itemize}
\item[(i)] A repeated randomness assumptions insuring that the off
diagonal elements of $\mbox{\boldmath $\rho$}_{i,j}$ vanish. This is
an assumption on the states that are accessible as the system
evolves.
\item[(ii)] The interaction potential that causes the mixing of states
is assumed to be weak as to permit the use of first order perturbation
theory.
\item[(iii)] There is an inherent clash in that the ${\cal P}_i$'s
refer to discrete states and yet one has to use continuum
normalization in order to obtain a Dirac delta function in the
energies.
\end{itemize} 
Conditions (i) and (ii) can be weakened for certain classes of
interactions \cite{Fujita} (also see L. Van Hove's in
Ref. \cite{Cohen}).

In this presentation we shall consider a nonisolated, finite but
large, system that is allowed to interact with its environment; by
this we mean that from time to time the system is exposed to a low
energy, low intensity packet of some quanta. These quanta are
spatially unconstrained (continuum normalized) and after a certain
characteristic interaction time $\tau$ the system settles into a new
state or density matrix.  After intervals larger than this
characteristic time, this process is repeated. The above qualitative
terms, low intensity, low energy and time interval, will be made
precise and summarized in Sec. \ref{conclusion}. The quanta in
these pulses will be referred to as ``photons''.  We shall show that
with one assumption on the interaction of these packets with the
system, {\em but independent of any assumptions on the accessibility
of the system itself}, microcanonical equilibrium,
eq.~(\ref{densitymatrix}), will be reached. Contrary to the approach
to equilibrium for isolated systems, in this case equilibrium is
attained by ``entangling'' the system with photons and then looking at
expectation values of operators that are diagonal in the photon
variables. Even though we may start out with a pure system-photon
state, the system itself will be in a mixed state after the
interactions have ceased. No recourse is made to perturbation theory;
the interactions are treated exactly.

The system, its interactions and time evolution are presented in
Sec. \ref{interactions} while the time evolution of density
matrices is discussed in Sec. \ref{t-evol}; an assumption on the
randomness of phases of certain amplitudes is presented and discussed
in this section. How the system approaches equilibrium is shown in 
\ref{Equilibrium}; although we do not get a master equation the
approach to equilibrium is Markovian \cite{Cohen}, governed by a
Chapman-Kolmogorov \cite{Cohen} equation. The relations among various
energy and time scales that the system and perturbing quanta have to
satisfy are given in Sec. \ref{conclusion} where the origin of time
irreversibility and the absence of recurrence is discussed. 

\section{Interactions and time evolution of states}\label{interactions} 
The energy levels $E_{\alpha}$ of the system are highly degenerate
with states $|\alpha, i\rangle$ .The energies of the photons in the
packets are taken to be too small to cause transitions between states
of different energy; these interactions can, however, cause
transitions between the degenerate levels within a given $E_{\alpha}$;
for convenience we take the energy $E_{\alpha}=0$ and drop the quantum
number $\alpha$ in the description of the states. As usual the total
Hamiltonian will be split into two parts, $H=H_0+H_1$; eigenstes of
$H_0$ that are of interest are $|i,{\mbox{\boldmath $k$}}\rangle$ with
\begin{equation}\label{freeH} 
H_0|i,{\mbox{\boldmath $k$}}\rangle =
\omega(k)|i,{\mbox{\boldmath
$k$}}\rangle\, ;
\end{equation} 
the specific dispersion relation for $\omega(k)$ is unimportant. $H_1$ 
induces transitions between $|i,{\mbox{\boldmath $k$}}\rangle
\leftrightarrow |j,\mbox{\boldmath $k$}'\rangle$. Whether this
Hamiltonian is time reversal invariant or not is immaterial to
subsequent developments.

Let $|i,{\mbox{\boldmath $k$}}\rangle_H$ be that eigenstate of the
total Hamiltonian that approaches $|i,{\mbox{\boldmath $k$}}\rangle$
for large negative times; we use the notation $|\ \rangle_H$ for an
eigenstate of the total Hamiltonian, whereas $|\ \rangle$, without the
subscript $H$, for an eigenstate of $H_0$. This state satisfies the
Lippmann-Schwinger equation \cite{GW}
\begin{equation}\label{L-S} |i,{\mbox{\boldmath
$k$}}\rangle_H=|i,{\mbox{\boldmath $k$}}\rangle
+\frac{1}{\omega(k)-H_0+i\epsilon}H_1|i,{\mbox{\boldmath
$k$}}\rangle_H\, .
\end{equation} The overlap with an eigenstate of the free
Hamiltonian is
\begin{equation}\label{overlap}
\langle j,\mbox{\boldmath $p$}|i,\mbox{\boldmath $k$}\rangle_H =
   \delta_{i,j}\delta({\mbox{\boldmath $k$}}-{\mbox{\boldmath $p$}})+
      \frac{1}{\omega(k)-\omega(p)+i\epsilon}
         {\cal T}_{j,i}(\mbox{\boldmath $p$},\mbox{\boldmath $k$})\, ,
\end{equation} where 
\begin{equation}\label{defT} {\cal T}_{j,i}(\mbox{\boldmath 
$p$},\mbox{\boldmath
$k$})=
\langle j,\mbox{\boldmath $p$}|H_1|i,\mbox{\boldmath $k$}\rangle_H 
\end{equation} is the scattering amplitude for the transition 
$|i,{\mbox{\boldmath $k$}}\rangle\to |j,{\mbox{\boldmath
$p$}}\rangle$and satisfies the off shell unitarity relations
\begin{eqnarray}\label{unitarity1} 
{\cal T}_{j,i}(\mbox{\boldmath
$p$},\mbox{\boldmath $k$})-
  {\cal T}^*_{i,j}(\mbox{\boldmath $k$},\mbox{\boldmath $p$})&=&
  \int d\mbox{\boldmath$q$}
   \left[\frac{1}{\omega(q)-\omega(p)+i\epsilon}-
      \frac{1}{\omega(q)-\omega(k)-i\epsilon}\right]
        {\cal T}_{j,n}(\mbox{\boldmath $p$},\mbox{\boldmath $q$})
       {\cal T}^*_{i,n}(\mbox{\boldmath $k$},\mbox{\boldmath $q$})\, ,
         \nonumber\\
    &=&\int d\mbox{\boldmath$q$}
   \left[\frac{1}{\omega(q)-\omega(p)+i\epsilon}-
      \frac{1}{\omega(q)-\omega(k)-i\epsilon}\right]
        {\cal T}_{n,i}(\mbox{\boldmath $q$},\mbox{\boldmath $k$})
       {\cal T}^*_{n,j}(\mbox{\boldmath $q$},\mbox{\boldmath $p$})\, .
\end{eqnarray}

We ant to describe the time evolution of a state that at $t=0$ 
started out
as $|i,\mbox{\boldmath
$k$}\rangle$; more specifically we want the overlap at time $t$ with
the state $|j,\mbox{\boldmath $p$}\rangle$.
\begin{equation}\label{timeevol1}
\langle j,\mbox{\boldmath $p$}|e^{-iHt}|i,{\mbox{\boldmath
  $k$}}\rangle = [\delta_{i,j}\delta(\mbox{\boldmath $p$}
     -\mbox{\boldmath $k$})+
       i{\cal R}_{j,i}(\mbox{\boldmath $p$},\mbox{\boldmath $k$};t)]
           e^{-i\omega(p)t}\, ;
\end{equation} in the above
\begin{eqnarray}\label{timeevol2} i{\cal R}_{j,i}(\mbox{\boldmath
$p$},\mbox{\boldmath $k$};t)&=&
 \frac{1}{\omega(p)-\omega(k)-i\epsilon}\left[
     -{\cal T}_{j,i}(\mbox{\boldmath $p$},\mbox{\boldmath $k$})
         e^{i\left[\omega(p)-\omega(k)\right]t}+
           {\cal T}^*_{i,j}(\mbox{\boldmath $k$},\mbox{\boldmath $p$})
                 \right]\nonumber\\
   &&+\int d\mbox{\boldmath$q$}e^{i\left[\omega(p)-\omega(q)\right]t}
   \frac{{\cal T}_{j,n}(\mbox{\boldmath $p$},\mbox{\boldmath $q$})}
      {\omega(q)-\omega(p)+i\epsilon}\, 
  \frac{{\cal T}^*_{i,n}(\mbox{\boldmath $k$},\mbox{\boldmath $q$})}
    {\omega(q)-\omega(k)-i\epsilon}\, .
\end{eqnarray} For the definition of ${\cal R}$ in eq.~(\ref{timeevol1}) 
we have
pulled out the factor $\exp\left[-i\omega(p)\right]$ for later  
convenience.
Eq.~(\ref{unitarity1}) or (\ref{timeevol1}) implies unitarity relations for
${\cal R}$ 
\begin{eqnarray}\label{unitarity2} -i\left[{\cal R}_{j,i}
(\mbox{\boldmath $k$}',\mbox{\boldmath $k$};t)
  -{\cal R}^*_{i,j}(\mbox{\boldmath $k$},\mbox{\boldmath $k$}';t)
   \right ]&=&
    \sum_n\int d\mbox{\boldmath$q$}
       {\cal R}_{j,n}(\mbox{\boldmath $k$}',\mbox{\boldmath $q$};t)
     {\cal R}^*_{i,n}(\mbox{\boldmath $k$},\mbox{\boldmath $q$};t)\, ,
\nonumber\\ &=&\sum_n\int d\mbox{\boldmath$q$}
       {\cal R}_{n,i}(\mbox{\boldmath $q$},\mbox{\boldmath $k$};t)
     {\cal R}^*_{n,j}(\mbox{\boldmath $q$},\mbox{\boldmath $k$}';t)\, .
\end{eqnarray}

\section{Time evolution of the density matrix}\label{t-evol}
We shall be interested in operators ${\cal O}$ be that have
off diagonal matrix elements between the different $|i\rangle$'s and
are diagonal in the photon subspace. For any state 
\begin{equation} 
|S\rangle=\sum_i\int d\mbox{\boldmath 
$q$}\phi_i(\mbox{\boldmath
$q$}) |i,\mbox{\boldmath $q$}\rangle\, ,
\end{equation} 
the expectation value of ${\cal O}$ is
\begin{equation}\label{entangle1}
\langle S|{\cal O}|S\rangle=
\sum_{i,j}\langle j|{\cal O}|i\rangle 
     \mbox{\boldmath $\rho$}_{i,j}\, ,
\end{equation} 
with a density matrix 
\begin{equation}\label{entangledensmatrix}
\mbox{\boldmath $\rho$}_{i,j}=\int d\mbox{\boldmath $q$}
\phi_i(\mbox{\boldmath $q$})\phi_j^*(\mbox{\boldmath $q$})\, .
\end{equation}

\subsection{Evolution of \mbox{\boldmath $\rho$} entangled with a wave
packet of photons} 
Suppose that at time $t=0$ our system, described by a density 
matrix $\mbox{\boldmath $\rho$}_{j,i}(0)$, is exposed to a photon state.
The total density matrix, for the system plus photons is
\begin{equation}\label{t=0densmatrix}
\mbox{\boldmath $\rho$}_T=\sum_{i,j} \int d\mbox{\boldmath $k$}
\, d\mbox{\boldmath $k$}' \psi(\mbox{\boldmath $k$})
\psi^*(\mbox{\boldmath $k$}') |j,\mbox{\boldmath $k$} \rangle 
\mbox{\boldmath $\rho$}_{j,i}(0)
\langle i,\mbox{\boldmath $k$}'|\, ,
\end{equation}  
where $\psi(k)$ describes the photon wave packet scattered of the mixed
state at $t=0$; $\int d\mbox{\boldmath $k$} |\psi(\mbox{\boldmath
$k$})|^2=1$. Using eq.~(\ref{timeevol1}) and eq.~(\ref{timeevol2}) the
density matrix at time $t$, after summing over the photon states, is
\begin{eqnarray}\label{t=tdensmatrix}
\mbox{\boldmath $\rho$}_{j,i}(t)=\mbox{\boldmath $\rho$}_{j,i}(0) &+&\int
d\mbox{\boldmath $k$}\, d\mbox{\boldmath $k$}'
\psi(\mbox{\boldmath $k$})\psi^*(\mbox{\boldmath $k$}')
\Big [i\sum_n{\cal R}_{j,n}(\mbox{\boldmath
$k$}',\mbox{\boldmath $k$};t) 
\mbox{\boldmath $\rho$}_{n,i}(0)\nonumber\\&-& i\sum_m{\cal
R}^*_{i,m}(\mbox{\boldmath $k$},\mbox{\boldmath $k$}';t)
\mbox{\boldmath $\rho$}_{jm}(0)+
\int d\mbox{\boldmath $p$}\sum_{n,m}  {\cal R}_{j,n}(\mbox{\boldmath
$p$},\mbox{\boldmath $k$};t) {\cal R}^*_{i,m}(\mbox{\boldmath
$p$},\mbox{\boldmath $k$}';t)
\mbox{\boldmath $\rho$}_{n,m}(0)
\Big ] .\nonumber\\
\end{eqnarray} 
In subsequent discussions we shall need some randomness condition on the
phases of the ${\cal R}$'s. It is unlikely that such a condition could
be valid for all times; we shall try for ones that may hold at large
times. For short duration pulses we expect that after some
characteristic collision time $\tau$, as for example the inverse of
the width of a resonance in resonance dominated scattering
\cite{t-3/2}, the system will settle down and
\begin{equation}\label{asympt1}
\lim_{t\to\infty}\int d\mbox{\boldmath $k$}\psi(\mbox{\boldmath $k$})
  {\cal R}_{j,i}(\mbox{\boldmath $p$},\mbox{\boldmath $k$};t)=
    {\cal R}_{j,i}(\mbox{\boldmath $p$})\, ;
\end{equation}
${\cal R}_{j,i}(\mbox{\boldmath $p$})$ depends implicitly on the wave 
function
$\psi$ and, using the definition in eq.~(\ref{timeevol2}) we find
\begin{equation}\label{asymptot1'} {\cal R}_{j,i}(\mbox{\boldmath $p$})=
\int d\mbox{\boldmath $k$}\psi(\mbox{\boldmath $k$})
\frac{-i}{\omega(p)-\omega(k)-i\epsilon}{\cal T}^*_{i,j}(
\mbox{\boldmath $k$},\mbox{\boldmath $p$})\, .
\end{equation} 
For further developments, this explicit form will not be needed. We
also define  
\begin{equation}\label{asymptot2} 
{\cal R}_{j,i}=\int d\mbox{\boldmath
$k$}\psi^*(\mbox{\boldmath $k$}')   {\cal R}_{j,i}(\mbox{\boldmath 
$k$}')\, .
\end{equation} 
These, in turn, satisfy the unitarity relations
\begin{equation}\label{unitarity3} 
-i\left({\cal R}_{j,i}-{\cal 
R}^*_{i,j}\right)=
 \sum_n\int d\mbox{\boldmath $p$}{\cal R}_{n,i}(\mbox{\boldmath $p$})
   {\cal R}^*_{n,j}(\mbox{\boldmath $p$})\, .
\end{equation} 
In particular, we find
\begin{equation}\label{positivity}
\mbox{\rm Im}\, {\cal R}_{j,j}>0\, .
\end{equation}
The evolution of the density matrix, eq.~(\ref{t=tdensmatrix}),  may
be expressed as  
\begin{equation}\label{t=tdensmatrix'}
\mbox{\boldmath $\rho$}_{j,i}(t)=\mbox{\boldmath $\rho$}_{j,i}(0)+
 i\sum_n\left[{\cal R}_{j,n}\mbox{\boldmath $\rho$}_{n,i}(0)-
  {\cal R}^*_{i,n}\mbox{\boldmath $\rho$}_{j,n}(0)\right] +\sum_{m,n}\int
d\mbox{\boldmath $p$}{\cal R}_{j,n}(\mbox{\boldmath $p$})
 {\cal R}^*_{i,m}(\mbox{\boldmath $p$})
    \mbox{\boldmath $\rho$}_{n,m}(0)\, .
\end{equation}

\subsection{Randomness Assumption}\label{Randomness Assumptions}
In order to proceed further we must impose a crucial condition on the
${\cal R}$'s: 
\begin{equation}\label{assumption}
\int d\mbox{\boldmath $p$} {\cal
R}_{j,n}(\mbox{\boldmath $p$}){\cal R}^*_{i,m}(\mbox{\boldmath $p$})
=0\ \  \mbox{for}\  i\ne j\ \  \mbox{and}\  m\ne n.
\end{equation} 
This results from the assumption that, as we integrate over
\mbox{\boldmath $p$}, the phases of ${\cal R}_{j,n}(\mbox{\boldmath
$p$})$ fluctuate rapidly. {\em It should be emphasized that this is an
assumptions on the dynamics of the system and not on states or density
matrices of the system at any particular time.}

The above and unitarity relation, eq.~(\ref{unitarity3}), lead to 
\begin{equation}\label{randomconseq1}
\left({\cal R}_{j,i}-{\cal R}^*_{i,j}\right)=0\,\ \ \mbox{\rm for}\ \  
i\ne j\, ;
\end{equation} 
which together with eq.~(\ref{positivity}) implies
\begin{equation}\label{Rdecomp}
{\cal R}={\cal R}_H+i{\cal D}
\end{equation}
with ${\cal R}_H$ Hermitian and ${\cal D}$ a diagonal matrix with
positive elements. As a matter of fact, there always exists a basis
of the states $|i\rangle$ where such a decomposition of ${\cal R}$
holds. Any matrix can be written as a sum of a Hermitian and an
anti-Hermitian one and we go to the basis where the anti-Hermitian part
is diagonal. Although eq.~(\ref{assumption}) implies
eq.~(\ref{Rdecomp}), the inverse is not true and eq.~(\ref{assumption})
remains an assumption. We work in the basis where this
decomposition holds. The time evolution of the density matrix becomes 
\begin{equation}\label{timeevol3}
\mbox{\boldmath $\rho$}_{j,i}(t)=\mbox{\boldmath $\rho$}_{j,i}(0)+  
  i\left[{\cal R}_H,\mbox{\boldmath $\rho$}(0)\right]_{j,i}-
    \left[{\cal D},\mbox{\boldmath $\rho$}(0)\right]_{j,i}
      +\delta_{i,j}\sum_n\int
d\mbox{\boldmath $p$}|{\cal R}_{i,n}(\mbox{\boldmath $p$})|^2
     \mbox{\boldmath $\rho$}_{n,n}(0)\, ;
\end{equation}
$[A,B]$ is the commutator of the matrices $A$ and $B$.

\section{Approach to Equilibrium}\label{Equilibrium}
At this point we have to require the ${\cal R}_{j,i}$'s to be small;
how this is achieved will be made precise in Sec.
\ref{conclusion}. We solve eq.~(\ref{timeevol3}) to first order in the
${\cal R}$'s by first rewriting it as 
\begin{equation}\label{timeevol4}
\mbox{\boldmath {$\tilde\rho$}}_{j,i}(t)=\mbox{\boldmath $\rho$}_{j,i}(0)
-\left[{\cal D},\mbox{\boldmath $\rho$}(0)\right]_{j,i}
      +\delta_{i,j}\sum_n\int
d\mbox{\boldmath $p$}|{\cal R}_{i,n}(\mbox{\boldmath $p$})|^2
     \mbox{\boldmath $\rho$}_{n,n}(0)\, ,
\end{equation}
with
\begin{equation}
\mbox{\boldmath {$\tilde\rho$}}_{j,i}(t)=
\sum_{n,m}\left[1-i{\cal R}\right]_{j,n}\mbox{\boldmath $\rho$}_{n,m}(t)
   \left[1+i{\cal R}\right]_{m,i}\, .
\end{equation}
To the order we are working $1-i{\cal R}$ is a unitary matrix and 
$\mbox{\boldmath
{$\tilde\rho$}}_{j,i}(t)$ is the density matrix in a basis rotated
from the one we started out at $t=0$. 

Let us first look at eq.~(\ref{timeevol4}) for $i\ne j$.
\begin{equation}
\mbox{\boldmath {$\tilde\rho$}}_{j,i}(t)=\left[\mbox{\bf $1$}
 -\left({\cal D}_{i,i}+{\cal D}_{j,j}\right)\right]
   \mbox{\boldmath $\rho$}_{j,i}(0)\, .
\end{equation}
The off diagonal elements of the unitarily rotated density matrix at
time $t$ are smaller than the corresponding matrix element at $t=0$;
\begin{equation}\label{t=ta}
|\mbox{\boldmath {$\tilde\rho$}}_{j,i}(t)|\le 
  |\mbox{\boldmath $\rho$}_{j,i}(0)|\, .
\end{equation}
For the case $i=j$ we have
\begin{equation}\label{t=tb}
\mbox{\boldmath {$\tilde\rho$}}_{i,i}(t)=
    \mbox{\boldmath $\rho$}_{i,i}(0)+i\left({\cal R}_{i,i}-
  {\cal R}^*_{i,i}\right)\mbox{\boldmath $\rho$}_{i,i}(0)+\sum_n\int
d\mbox{\boldmath $p$}|{\cal R}_{i,n}(\mbox{\boldmath $p$})|^2
     \mbox{\boldmath $\rho$}_{n,n}(0)\, .
\end{equation}
The coefficient of $\mbox{\boldmath $\rho$}_{n,n}(0)$ in the last term of
this equation may be identified with the transition probability for
$|n\rangle$ to evolve into $|i\rangle$,
\begin{equation}\label{transprob1}
{\cal W}_{i\leftarrow n}=\int d\mbox{\boldmath $p$} 
   |{\cal R}_{i,n}(\mbox{\boldmath $p$})|^2\, .
\end{equation}
Using the unitarity relation, eq.~(\ref{unitarity3}), we find 
\begin{equation}
i({\cal R}_{i,i}-{\cal R}^*_{i,i})=-\sum_n{\cal W}_{n\leftarrow i}
\end{equation}
and the evolution equation for this case becomes
\begin{equation}\label{CK} 
\mbox{\boldmath {$\tilde\rho$}}_{i,i}(t)=
\mbox{\boldmath $\rho$}_{i,i}(0)-\sum_n W_{n\leftarrow i}
\mbox{\boldmath $\rho$}_{i,i}(0)
  +\sum_n W_{i\leftarrow n}\mbox{\boldmath $\rho$}_{i,i}(0)\, .
\end{equation} 
Even if we include the unitary rotations this equation is of the
Chapman-Kolmogorov type \cite{Cohen} and for the ${\cal
W}_{i\leftarrow n}$'s not too large, will directly yield
microcanonical equilibrium.  This follows from the observation that
the matrix
\begin{equation} 
M_{i,j}=-\delta_{i,j}\sum_n {\cal W}_{n\leftarrow i} 
+{\cal W}_{i,j}
\end{equation} 
has one eigenvalue equal to zero and all others are negative which in
turn is obtained by showing the function $H(\tau)=-\sum_i{\cal P}_i(\tau)
\ln{\cal P}_i(\tau)$ satisfies a Boltzmann H-theorem, $dH/d\tau\ge 0$
with the probabilities being functions of the auxiliary variable
$\tau$ and (c.f. eq~(\ref{Pauli}))
\begin{equation}
\frac{d{\cal P}_i}{d\tau}=\sum_j M_{i,j}{\cal P}_j\, .
\end{equation} 
Let $v^{(0)}_i\sim (1,1,\cdots,1 )$ be the eigenvector of $M_{i,j}$
with eigenvalue zero and $v^{(\alpha)}_i$ be all the others. 
\begin{equation}
\mbox{\boldmath $\rho$}_{i,i}=c_0v^{(0)}_i
+\sum_{\alpha}c_{\alpha}v^{(\alpha)}_i\, .
\end{equation}
At the end of the interval $t$ $c_0$ hasn't changed and the other
$c_{\alpha}$'s have all decreased in magnitude. We find that repeated
pulses will drive off diagonal elements to zero and the diagonal ones
to the same constant or the final density matrix will be as in
eq.~(\ref{densitymatrix}).  

\section{Concluding Remarks}\label{conclusion}
We have to consider three characteristic energies: (i)
$\Delta_\alpha$, the difference in the energy levels $E_\alpha$ of
the system, (ii) $\delta_\omega$, the energy spread of the impinging
wave packets, and (iii) $\gamma$, the inverse of the interaction time
$\tau$. These quantities have to satisfy
\begin{equation}\label{characenergies}
\Delta_\alpha\gg\delta_\omega\gg\gamma\, .
\end{equation}
The first of these inequalities insures that there will be no
transitions between the different energy levels of the system; for
such transition the energy denominator in eq.~(\ref{overlap}) 
would have a typical magnitude of $\Delta_\alpha$ as opposed to
$\delta_\omega$ for intra level transitions. Due to the second
inequality the ``arrival period'' of the pulse is much shorter than
the interaction time and for times greater than $\tau$ the system and
the photons propagate separately. In addition the repetition time of
these pulses, $T$, must satisfy 
\begin{equation}\label{charactimes}
T\gg\frac{1}{\gamma}\, ;
\end{equation} 
this guarantees that as a new packet interacts with the system it is
unencumbered by photons from the previous pulses; for example $T$
must be large enough to allow any possible system-photon resonances to
decay.

Till now, all this has been worked out for the system interacting
with a {\em one} photon packet; the random impulses from the outside 
are likely to contain many photons and we extend our results to
$N$ photons where the states are $\int\prod_{n=1}^N d\mbox{\bf k}_n\, 
\psi_n(\mbox{\bf k}_n)|i;\mbox{\bf k}_1,\mbox{\bf k}_2\cdots 
\mbox{\bf k}_N\rangle$. In the case the packets $\psi_n({\bf k})$ 
are different for different $n$'s this expression should be 
appropriately symmetrized; doing this explicitly would just lead 
to unnecessary notational complications. The scattering amplitude 
is not necessarily a sum of two body amplitudes and 
the rest of the development is as the previous one. We identify a
weak, or low intensity pulse as one with few photons and an intense
one with a large number. We require that $N$ be sufficiently small
that the ${\cal R}_{j,i}$'s are small compared to one; $N$ should 
not be too small as the time to reach equilibrium would become very 
large.
  
With these conditions and the assumption about interactions made in
eq.~(\ref{assumption}) satisfied, repeats interactions of the system
with external quanta will drive it to equilibrium. 
Time irreversibility is built in right at the start in the choice of
the sign of the $i\epsilon$ term in eq.~(\ref{L-S}) The
continuous energy levels of the photons preclude any recurrences; had
the photons also been quantized in a finite volume the limit
considered in eq.~(\ref{asympt1}) would not have existed.

\nobreak 


\begin{references}
\bibitem{Lebowitz}  J.L. Lebowitz, in {\it 25 Years of Non-Equilibrium
Statistical Mechanics, Proceedings, Sitges Conference, Barcelona, Spain, 
1994}, edited by J.J. Brey {\it et al\/.} (Springer, 1995). 
\bibitem{chaos} N.G. Van Kampen in {\it Chaotic Behavior in Quantum 
Systems}, edited by G. Casati (Plenum Press, New York, 1985);
M. Srednicki, Phys.\ Rev.\ E{\bf 50}, 888 (1994). 
\bibitem{Cohen}  L. Van Hove in {\it Fundamental Problems in   
Statistical Mechanics}, edited by E.D.G. Cohen (North Holland,
Amsterdam, 1962), p. 157; N.G. Van Kampen, {\it ibid\/.}, p. 173.
\bibitem{Fujita} S. Fujita, {\it Introduction to Non-Equilibrium
Quantum Statistical Mechanics} (W.B. Saunders Company, Philadelphia,
1966).
\bibitem{GW} M.L. Goldberger and  K.M.~Watson, {\it Collision Theory} 
(John Wiley \& Sons, Inc, New York, 1964).
\bibitem{t-3/2} We ignore the very small nonexponential terms that
behave as $t^{-\frac{3}{2}}$; see Ref.~\cite{GW}.
\end{references}
\end{document}